\documentclass[reprint, superscriptaddress, amsmath, amssymb, aps, prx, longbibliography]{revtex4-1}
\usepackage{color}
\usepackage{hyperref}
\usepackage{graphicx}	
\usepackage{dcolumn}   	
\usepackage{bm}		    
\usepackage{epstopdf}
\usepackage{float}

\usepackage{array}
\newcolumntype{C}[1]{>{\centering\let\newline\\\arraybackslash\hspace{0pt}}m{#1}}

\begin{document}
\title{Dynamics of spin polarization in tilted polariton rings}

\author{S. Mukherjee}
\affiliation{Department of Physics and Astronomy, University of Pittsburgh, Pittsburgh, PA 15260, USA}
\author{V. K. Kozin}
\affiliation{Science Institute, University of Iceland, Dunhagi-3, IS-107 Reykjavik, Iceland}
\affiliation{ITMO University, St. Petersburg 197101, Russia}
\author{A. V. Nalitov}
\affiliation{ITMO University, St. Petersburg 197101, Russia}
\affiliation{Faculty of Science and Engineering, University of Wolverhampton, Wulfruna St, Wolverhampton WV1 1LY, UK}
\affiliation{Institut Pascal, PHOTON-N2, Universit\'e Clermont Auvergne, CNRS, SIGMA Clermont, Institut Pascal, F-63000 Clermont-Ferrand, France.}
\author{I. A. Shelykh}
\affiliation{Science Institute, University of Iceland, Dunhagi-3, IS-107 Reykjavik, Iceland}
\affiliation{ITMO University, St. Petersburg 197101, Russia}
\author{Z. Sun}
\affiliation{Department of Physics and Astronomy, University of Pittsburgh, Pittsburgh, PA 15260, USA}
\affiliation{State Key Laboratory of Precision Spectroscopy, East China Normal University, Shanghai 200062, China}
\author{D. M. Myers}
\affiliation{Department of Physics and Astronomy, University of Pittsburgh, Pittsburgh, PA 15260, USA}
\author{B. Ozden}
\affiliation{Department of Physics and Astronomy, University of Pittsburgh, Pittsburgh, PA 15260, USA}
\affiliation{Department of Physics and Engineering, Penn State Abington, Abington, PA 19001, USA }
\author{J. Beaumariage}
\author{M. Steger}
\affiliation{Department of Physics and Astronomy, University of Pittsburgh, Pittsburgh, PA 15260, USA}
\author{L. N. Pfeiffer}
\author{K. West}
\affiliation{Department of Electrical Engineering, Princeton University, Princeton, NJ 08544, USA}
\author{D. W. Snoke}
\affiliation{Department of Physics and Astronomy, University of Pittsburgh, Pittsburgh, PA 15260, USA}

\begin{abstract}
We have observed the effect of pseudo magnetic field originating from the polaritonic analog of spin-orbit coupling (TE$-$TM splitting) on a polariton condensate in a ring shaped microcavity. The effect gives rise to a stable four-leaf pattern around the ring as seen from the linear polarization measurements of the condensate photoluminescence. This pattern is found to originate from the interplay of the cavity potential, energy relaxation and TE-TM splitting in the ring. Our observations are compared to the dissipative one-dimensional spinor Gross-Pitaevskii equation with the TE-TM splitting energy which shows good qualitative agreement.   
\end{abstract}

\maketitle
\section{Introduction}

Over a decade has passed since the first demonstration of Bose-Einstein condensation of exciton-polaritons in a semiconductor microcavity \cite{balili2007bose,kasprzak2006bose}. It stands out as a unique platform for exploring physics of out of equilibrium systems. Recent reports on the observation of thermal equilibration \cite{mitprl,Caputo2017} on one hand have triggered interest in studying equilibrium physics in non-hermitian systems, while on the other have shown promise for making devices based on polariton condensates which operate near equilibrium. Such a feat been achieved in microcavities with Q $>10^5$, allowing several collision events between polaritons to distribute energy and thermalize in their lifetime \cite{mitprl,Caputo2017,steger-turn,steger2013long}.

These systems also naturally offer a way for describing a pseudo-spin 1/2 Bose gas \cite{Microcavities2017}. The order parameter of a polariton condensate is described by a two-component complex valued spinor, which is connected to the electric polarization of the polaritons in the microcavity \cite{snoke2020solid}. In the presence of a non-aligned magnetic field the spin of the polaritons is precessing similarly to the spin of an electron in a magnetic field. It turns out that for a quantum well embedded inside a semiconductor microcavity, a small but non-negligible momentum dependent effective magnetic field is present which lies in the plane of the quantum well. This has been used to realize a polaritonic analog of extrinsic \cite{leyder2007observation} as well as intrinsic \cite{kammann2012nonlinear} spin Hall effect. In contrast to the electronic systems where the spin Hall effect gives rise to a spin current and no mass transport, the optical spin Hall effect is realized by actual transport of polaritons while the spin of the polaritons precess as they move.    

In this paper we address how the two spinor components of a gas of highly excited trapped polaritons evolve following a quench in a long lifetime ($\approx 200$ ps) microcavity. The trapping potential is created by patterning the top mirror of the GaAs microcavity in the shape of a ring. The polaritons in the ring maintain the same long lifetime ($\approx$ 200 ps), high quality (Q $> 10^{5}$) and low disorder as in the two-dimensional planar microcavity used previously \cite{pnas,steger-turn,Steger-OL,mitprl,Myers-apl}. With the presence of a unidirectional gradient in the energy of the polaritons, the circular symmetry of the ring is broken, giving rise to a rigid rotor potential. Combined with a momentum dependent effective magnetic field in the trap, this results in intrinsic optical spin Hall effect as the polaritons are transported to the region with the lowest energy in the trap. We map the spin of the condensed polaritons in the ring using space- and time- resolved spin-polarized spectroscopy. 

The rings are fabricated by etching the top distributed Bragg reflector (DBR) of this microcavity in the shape of rings of width (the difference of the outer and the inner radii) 15 $\mu$m and radius (average of outer and inner radii) 50 $\mu$m. Further details may be found in the previous papers \cite{Myers-apl,MyersArxiv2018,mukherjee2019observation}. Across the typical dimensions of the ring, the thickness of the microcavity varies which leads to a cavity energy gradient ($\approx$ 7$-$9 meV/mm). The cavity gradient effect is similar to an artificial gravity for the polaritons making the rings look tilted on the potential energy plane. Due to the analogy to gravity, here we will refer to the point of highest potential energy as the ``top'' of the ring, and the point of lower potential energy, on the opposite side of the ring, as the ``bottom''. The effect of the artificial gravity on the spin-polarized polaritons have been theoretically studied in Ref. \cite{sedov2017artificial}.

\begin{figure*}
\centering
\includegraphics[width = 0.9\linewidth]{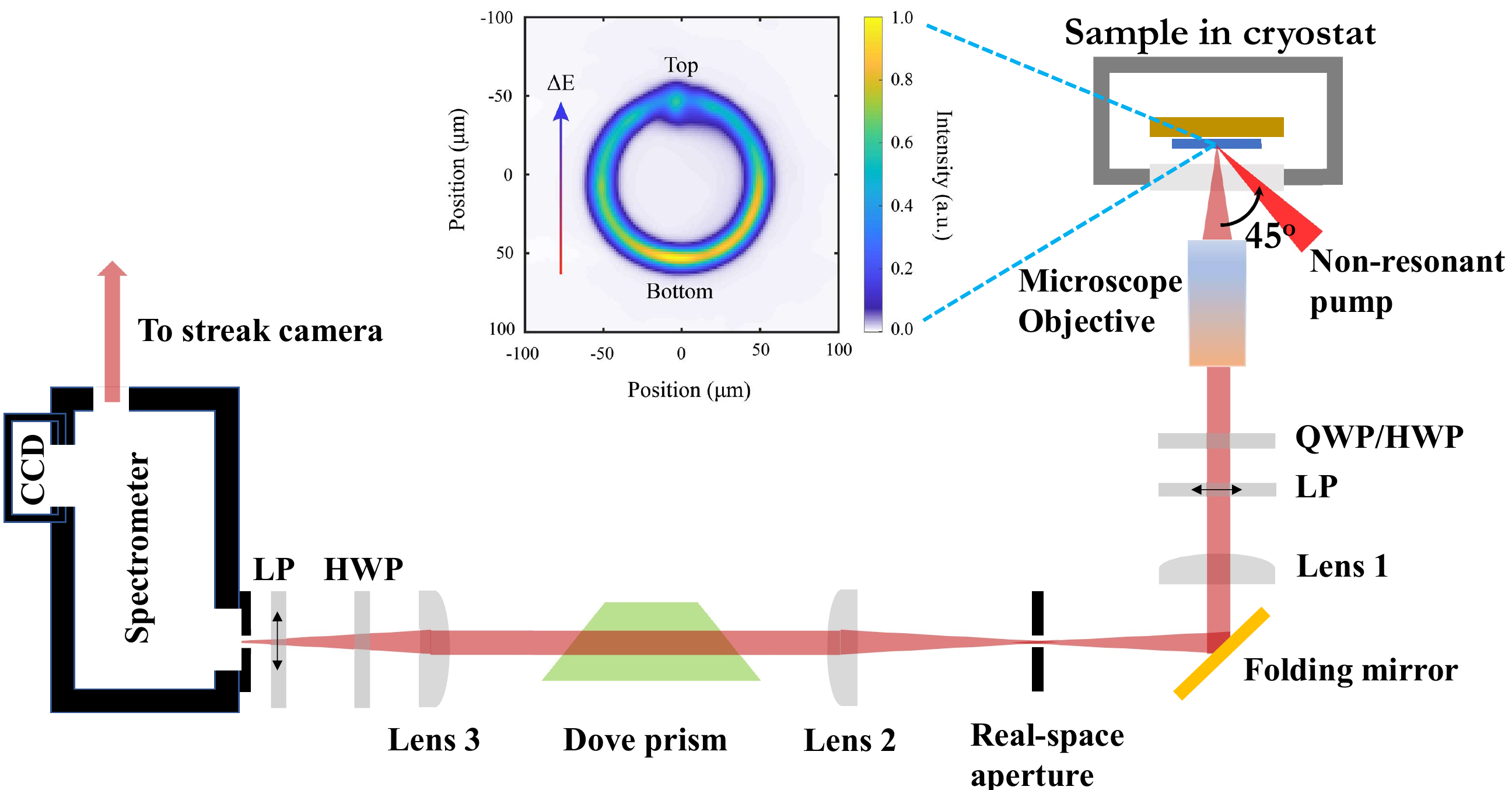}
\caption{Schematic of the experimental setup. Inset shows a time integrated image of the photoluminiscence from the ring microcavity also showing the direction of the cavity gradient $\Delta$E. QWP: quarter waveplate, HWP: half waveplate, LP: linear polarizer. } 
\label{fig:Fig1}
\end{figure*}





\section{Experiment}\label{Sec2}

The top of the rings are non-resonantly pumped ($E_\mathrm{pump} \approx 1710$ meV which is at least 100 meV higher than the energy of the lower branch polaritons at the point of excitation) with a mode-locked Ti:sapphire laser with a pulse repetition rate of 76 MHz, a pulse width of $\approx 2$ ps, a spot size of $\approx 15$ $\mu$m on the sample and incident at an angle of $\approx 45^\circ$ from the plane of the sample with more than 90$\%$ linear polarization. Due to the high angle injection, the pump spot is not circular on the plane of the sample and creates asymmetry in the direction of the polaritons streaming from the pump spot. Photoluminescence (PL) from the rings was collected using a microscope objective with a numerical aperture (NA) of 0.40 and imaged onto the entrance slit of a spectrometer. The image was then sent through the spectrometer either to a standard charged coupled device (CCD) chip located at one of the exit ports of the spectrometer for time integrated imaging, or onto a Hamamatsu streak camera located at the other exit port for time-resolved imaging. All measurements were performed by cooling the microcavity to low temperature (below 10 K) in a continuous-flow cold-finger cryostat. A sketch of the experimental setup is shown in Fig. \ref{fig:Fig1}. 

\begin{figure*}
\centering
\includegraphics[width = \linewidth]{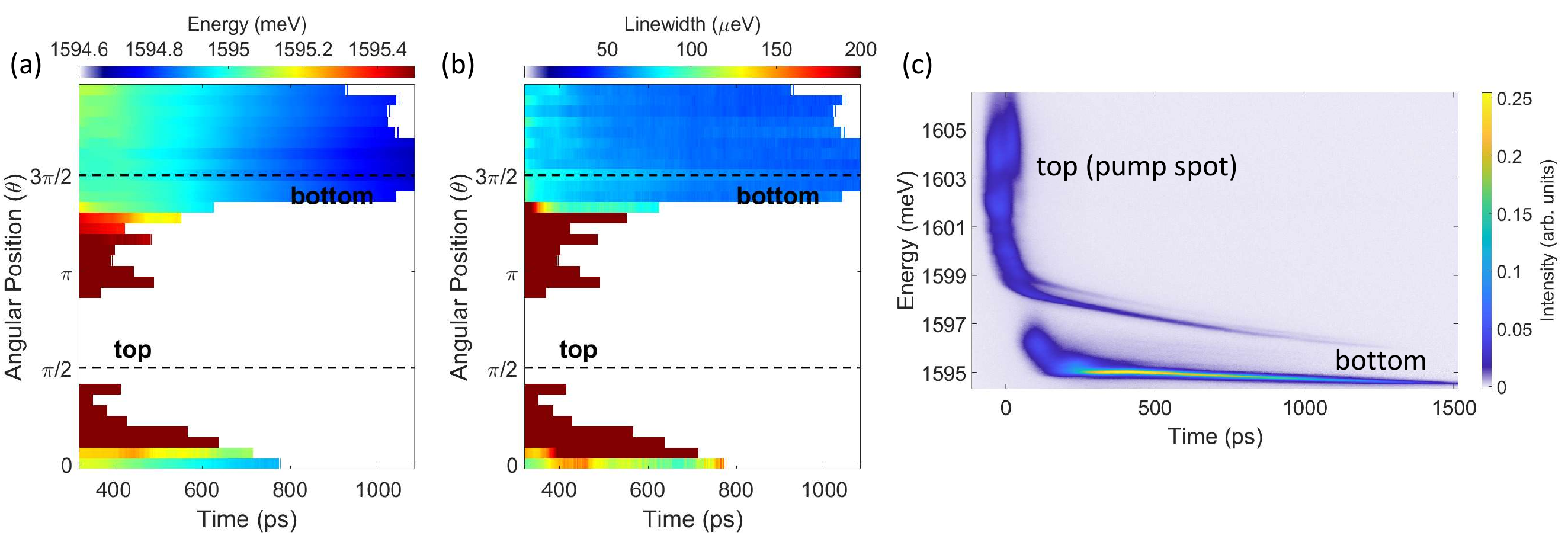}
\caption{Example of condensate energy (a) and linewidth (b) as a function of time from different locations on the ring. $3\pi/2$ corresponds to the bottom of the ring. The zero of the time axis corresponds to the time of excitation at the top of the ring. The condensate energy is obtained at each spatial location by fitting a lorentzian function to the energy- and time- resolved image captured by the streak camera. (c) shows a spectral comparison between the PL emission from the top and bottom of the ring as a function of time.  } 
\label{fig:Fig2}
\end{figure*}

The non-resonant optical pulse excites a population of free electrons and holes at the top of the ring which undergo rapid thermalization, turning into excitons. Excitons further relax down in energy and reach the anti-crossing spectral region of the photonic and excitonic dispersion branches, forming a dense polariton gas (see, for example, Tassone et al. \cite{tassone97}), which above a critical density undergoes non-equilibrium Bose-Einstein condensation. This non-equilibrium condensate streams out ballistically from the excitation spot and fills the entire ring, while also dropping in energy as it moves in the ring. As shown in Fig. \ref{fig:Fig2} (c) the polaritons are initially formed with a large blue-shifted energy at the pump spot which rapidly red shifts with time. This is followed by a spectrally narrow emission which slowly drops in energy, asymptotically approaching the $k_{||}=0$ lower polariton energy at this location. The initial blue shift seen at the pump spot is due to the interaction of polaritons with the exciton cloud. The rate of energy drop of the polaritons measures the rate at which the exciton cloud decays. By the time the polaritons arrive at the bottom of the ring trap it has already undergone considerable energy relaxation ($\approx$ 2 meV). The energy- and time- resolved image from the bottom in Fig. \ref{fig:Fig2} (c) shows that as the occupation density of the polaritons increases, the spectral linewidth of the emission narrows indicating build up of coherence in the bottom of the trap. In Fig. \ref{fig:Fig2} (a) and (b) we show spatially and temporally resolved energy and linewidth of the emission from the bottom of the trap. From these plots we infer that the spatial coherence extends at most to one-third of the ring circumference at any given point in time, which is signaled by the energy locking of the emission. At late times the energy of the emission approaches the local, low density $k_{||}=0$ lower polariton energy.     

We resolved the polarization of the PL from the ring by performing a full Stokes vector measurement ($S_0$, $S_1$, $S_2$, $S_3$) using combinations of half or quarter waveplate and a linear polarizer. The measurement was done on the collimated signal just after the microscope objective. The transmission axis of the polarizer was chosen to be vertical in the lab frame and was kept fixed throughout the experiment. This was done to remove the polarization sensitivity of the optics downstream in the setup. We used another half waveplate and a polarizer before the entrance slit of the spectrometer to collect all the signal which didn't get reflected, scattered or absorbed after passing through the optics. For all the measurements the orientation of the transmission axis of the final polarizer was also kept fixed, which removed the polarization sensitivity of all the optics inside the spectrometer as the light that entered the spectrometer was always at a fixed linear polarization. Two images were taken for each individual measurement, one with the half waveplate fast axis (placed just before the final polarizer) at 0$^{\circ}$ and one with it at 45$^{\circ}$. By adding these two images together, the total contribution of both polarizations (both parallel and orthogonal to the final polarizer) were taken into account. Additional details on the full Stokes vector measurement are given in Appendix \ref{ApdxA}.

Time-resolved Stokes vector measurements were done using the streak camera. A diametric slice of the image of the ring was aligned with the horizontal time slit of the streak camera. To collect the PL from a different location on the ring, the image was rotated using a dove prism. The dove prism was placed in a nearly collimated region of the optical path, as shown in Fig. \ref{fig:Fig1}. Before the experiment, the dove prism was carefully aligned to minimize the image walk-off when the prism was rotated by adjusting the tilt screws on the mount. Small adjustments were made during data collection by moving the final imaging lens, Lens 3, in the transverse plane to ensure consistent overlap of the diametric slice of the ring with the time slit. We note that the intensity collected by the streak camera after a pulsed excitation of the ring is the sum of the intensity of millions of such realizations. The density of polaritons created by each pulse was above the critical density of polaritons required for undergoing Bose-Einstein condensation (BEC) at the given temperature. If the realization of each instance of the BEC picks up a random polarization state in the ring, then by averaging we should obtain a strong component of unpolarized light in the emission. However, from previous time resolved studies on spontaneously formed polariton BEC it is known that the polarization of the BEC is not random and is sensitive to underlying crystal symmetries and other symmetries or imperfections of the microcavity. Therefore, degree of polarization is an important parameter distinguishing between emission from condensed and uncondensed or excited polaritons.   

\begin{figure}[H]
\centering
\includegraphics[width = 0.5\textwidth]{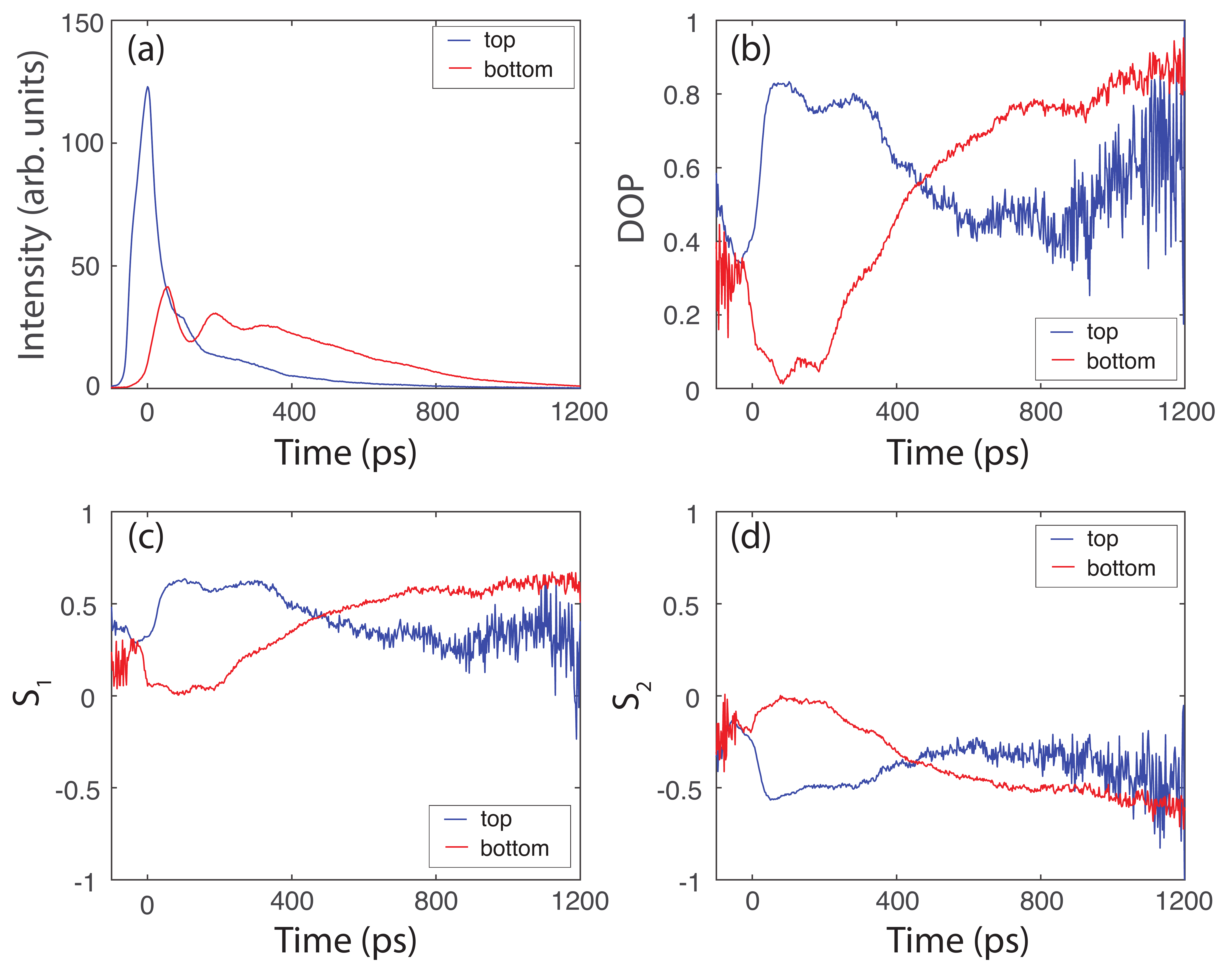}
\caption{Comparison of the polarization between the top and bottom of the ring as a function of time. Degree of polarization (DOP) is given by $\sqrt{S_1^2+S_2^2+S_3^2}$. The circular polarization $S_3$ is small and is not shown here.  }
\label{fig:Fig3}
\end{figure}

We compare the time resolved polarization of the PL emission from the point of excitation (top) and the diametrically opposite point (bottom) on the ring in Fig. \ref{fig:Fig3}. We find a rapid build up of the degree of polarization (DOP) of the emission at the pump spot following the arrival of the pump pulse. This is correlated with a decrease in the PL emission intensity at the pump spot. This signal is post hot thermalization and indicates appearance of a local non-equilibrium condensate which undergoes further energy relaxation. As the condensate streams and fills the ring, the DOP plateaus. At the same time, there is a slow build up of the polarized emission from the bottom of the ring which persists as long as the polaritons leak from the microcavity. In contrast, the DOP drops at the pump location after plateauing because the condensate drifts away from this location towards the bottom of the ring. The oscillations in the intensity of the PL emission from the bottom of the ring is due to the pendulum like oscillations about the ring trap minima \cite{mukherjee2019observation}.       

Measurement of the time-resolved components of the Stokes vector from different angular positions on the ring are shown in Fig. \ref{fig:Fig4} and a Bloch vector visualization at different time instances are shown in Fig. \ref{fig:Fig5}. The circular polarization component $S_3$ in these measurements reveal that before the appearance of condensate in the ring, highly excited non-thermal polaritons moving out from the pump spot show spin imbalance polarity depending on whether they are moving in the clockwise or anti-clockwise direction. This is due to the polariton pseudo-spins experiencing the effect of the in-plane magnetic field arising due to the splitting of the transverse electric (TE) and transverse magnetic (TM) mode in the cavity and depending on the polariton wave-vector. It is easily verified in a simulation showing the time evolution of a linearly polarized wavepacket under the TE-TM splitting Hamiltonian.     

The linear polarization components $S_1$ and $S_2$ show a spontaneous emergence of four-leaf angular pattern as shown in Fig. \ref{fig:Fig4} (b) and (c). This pattern persists in the ring until the polaritons fully leak from the microcavity. Manifestation of this pattern could be visualized by tracing at the projection of the Stokes vector on the equatorial plane of the Bloch sphere which gives the component of linear polarization of the emission. This is shown in black arrows in Fig. \ref{fig:Fig5} at different time instances. We see in these figures that after t = 265 ps, the linear polarization component appears to wrap around the ring by $4\pi$ radians, which corresponds to a rotation by $2\pi$ radians for the major axis of the elliptically polarized emission from the ring. The direction of the linear polarization component from the lower half of the ring remains nearly stationary in time after t = 265 ps. This is despite the fact that there is noticeable angular variation of the intensity of emission from this region with time as shown in $S_0$ component in Fig. \ref{fig:Fig4} (a). Intensity variation indicates density variation of the polaritons, from which we can conclude that the interactions between the polaritons do not play a dominant role in determining the polarization direction of the emission. This non-equilibrium state shows very little spin imbalance as shown in Fig. \ref{fig:Fig4} (d) which is due to good spatial overlap of nearly identical density profiles of the two spin components.   
	\begin{figure}[H]
		\centering
		\includegraphics[width=0.5\textwidth]{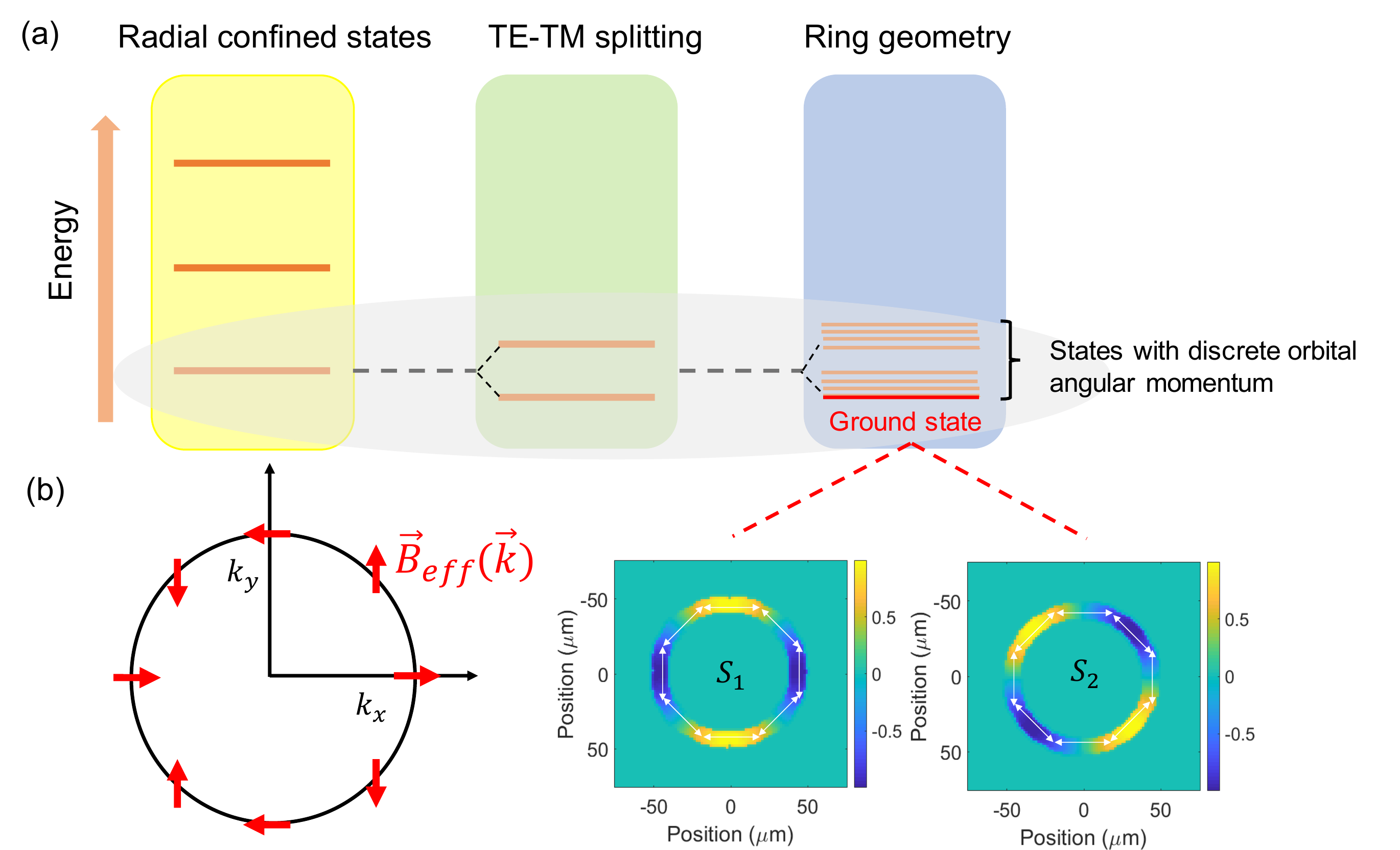}
		\caption{(a) A schematic showing the effect of the radial confinement, TE-TM splitting and the ring geometry on the quantum states in the ring. Visualization of the expectation value of the Stokes components in the ground state of the ring without the tilt is also shown. In the ground state $S_z=0$ while $S_1$ and $S_2$ components are phase shifted by $\pi/4$ w.r.t. each other. The white arrows depict the polarization plane direction, corresponding to the local Stokes parameters. In the ground state the polarization wraps in the azimuthal direction in the ring. (b) Direction of the effective magnetic field due to TE-TM splitting shown with red arrows in the momentum space. }
		\label{fig:Fig6}
	\end{figure}

\begin{figure*}
\centering
\includegraphics[width = \linewidth]{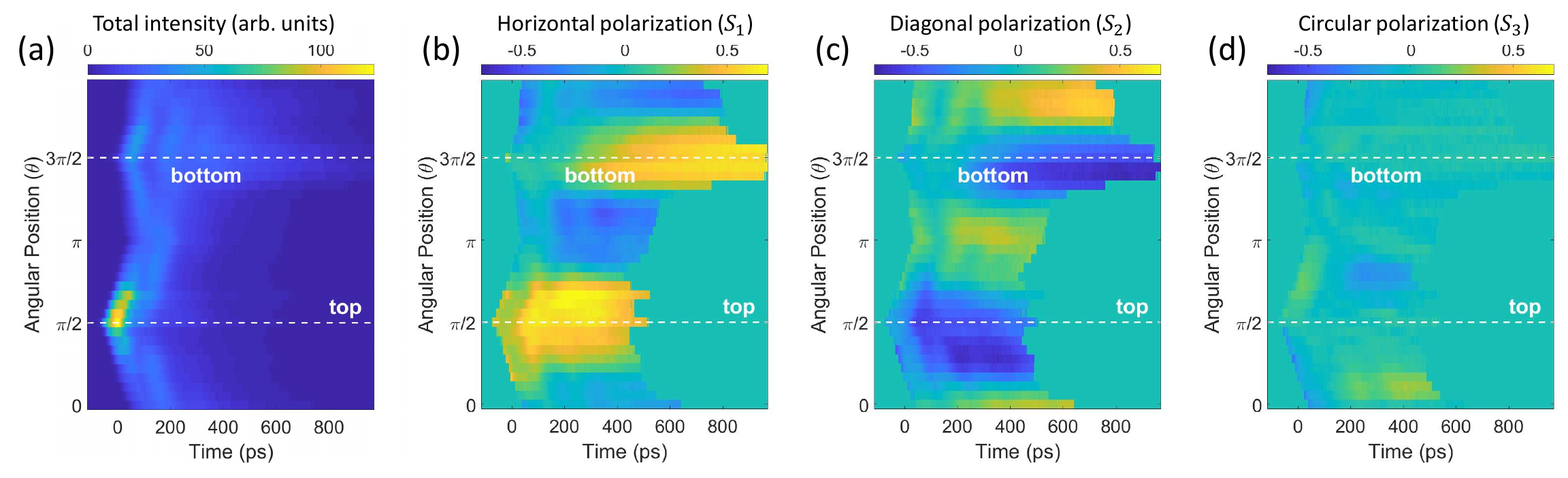}
\caption{Example of time and polarization resolved PL from tilted ring. Components of the Stokes vector $S_0, S_1, S_2$ and $S_3$ measured at an angular resolution of 10$^\circ$ as a function of time. We integrated the polarization resolved PL over the radial width of the ring to process the Stokes components at any given angular position on the ring. The intensity collected by the streak camera after a pulsed excitation of the ring is the sum of the intensity of millions of such realizations.} 
\label{fig:Fig4}
\end{figure*}

\begin{figure*}
\centering
\includegraphics[width = \linewidth]{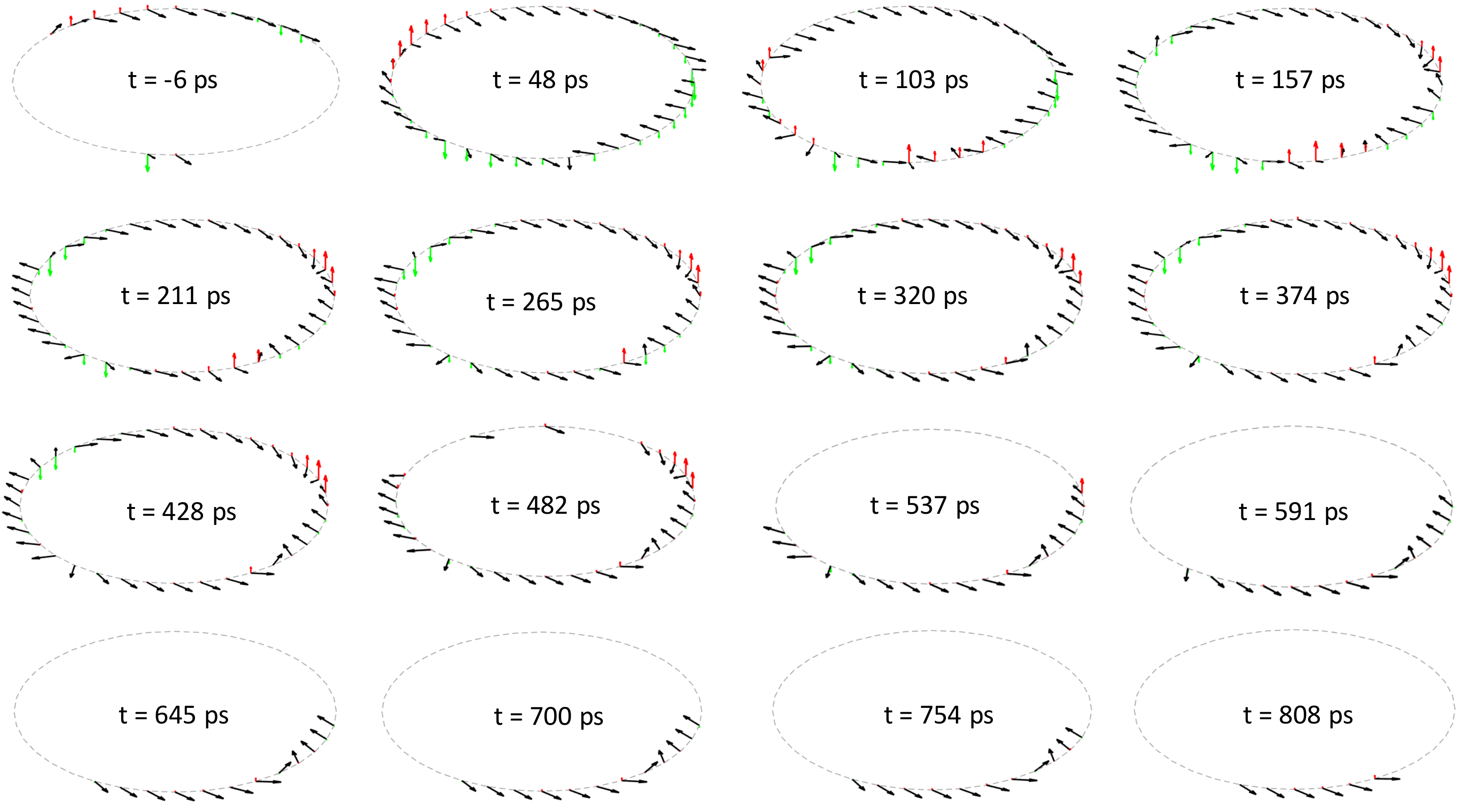}
\caption{Visualization of the components of the Stokes vector at different angular positions on the ring at different time instances. Black arrows depict the projection of the Stokes vector on the equatorial plane of the Bloch sphere. The length of the arrow is a measure of the degree of linear polarization and half of the angle w.r.t. the positive x-axis measures the direction of the linear polarized light. The red and green arrows depict the spin up and spin down projection of the polariton spinor respectively. The length of these arrows indicate the degree of circular polarization (or spin polarization) of the polaritons.} 
\label{fig:Fig5}
\end{figure*}

In the following section we develop a minimal model capturing the qualitative features discussed above highlighting the essential physics needed to interpret our observations. We derive an effective one-dimensional Hamiltonian by projecting the two-dimensional Hamiltonian onto the ground state of the radially confining potential due to the finite width of the ring channel. The periodicity of the wavefunction in the azimuthal direction results in discrete orbital angular momentum states in the ring. We assume that the polaritons occupy only a small part of the lower polariton dispersion near $k_{||}=0$, allowing us to make an effective mass approximation for the polaritons. The cavity gradient in the ring reduces the circular symmetry to a left-right mirror symmetry in the ring. It also creates a small spatial anisotropy in the TE-TM splitting energy which is ignored in the present model. A schematic of the quantum states in the ring in absence of a cavity gradient is shown in Fig. \ref{fig:Fig6} (a). We also neglect the spin-dependent interactions for reasons discussed previously.

\begin{figure*} \centering
\includegraphics[width = \linewidth]{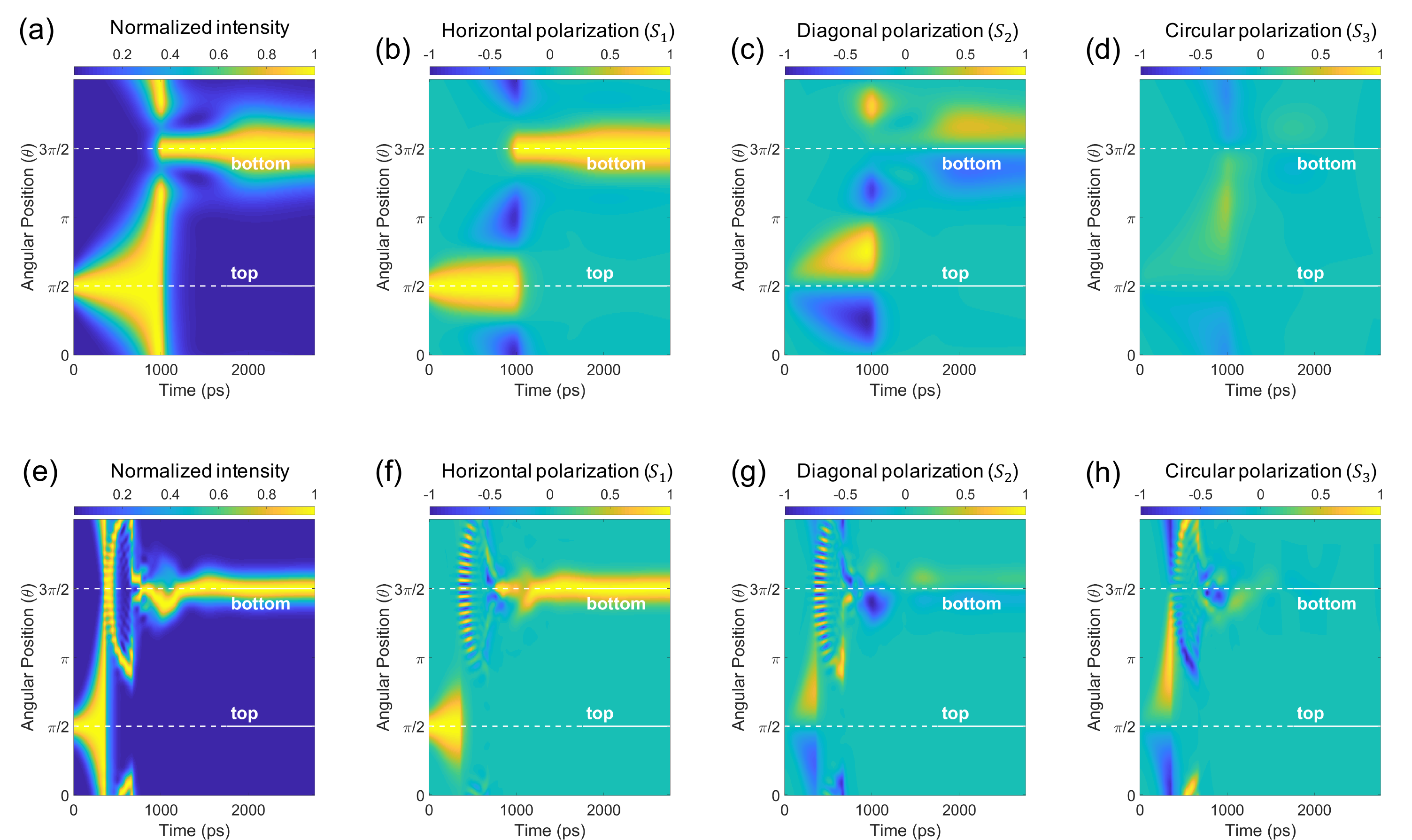}
\caption{Time evolution of the horizontally polarized condensate spinor. Parameters for both rows are $m_{eff}=6.14\times10^{-5}m_0$, $B=0.05$ and $\Delta = 250$, where $m_0$ is the rest mass of an electron in vacuum. Top row (a-d): $V_0$ = 16, $\Lambda$ = 1; Bottom row (e-h): $V_0$ = 335, $\Lambda$ = 0.3; A larger energy damping rate is chosen for smaller tilt case (top row) because the initial state has a lower potential energy than the larger tilt case (bottom row), so energy relaxation is slower for the same value of $\Lambda$ in the smaller tilt case. 
\label{fig:simulation}}
\end{figure*}

\section{Theoretical model}
Here we present the theoretical model of a single polariton ring of radius $R$ and width $a$ with a cavity gradient along the vertical axis (see the inset in Fig. \ref{fig:Fig1}), where only the lowest radial mode is occupied. In the absence of the polariton-polariton and polariton-reservoir interaction, the ring is described by the following matrix Hamiltonian $\widehat{H}$, whose elements are given by (see the derivation in Appendix~\ref{Appendix_derivation})

\begin{align}\label{eq:ham_matr}
&\widehat{H}_{11}=\widehat{H}_{22}=\frac{\hbar^2}{2m_{\text{eff}}R^2}(\hat{\tilde{k}}^2+V_0\sin{\varphi}),\\
&\widehat{H}_{12}=\widehat{H}^{\dagger}_{21}=\frac{\hbar^2}{2m_{\text{eff}}R^2}\cdot\nonumber\\
&\bigg((B+V_1\sin{\varphi}) e^{-2i\varphi}(-\hat{\tilde{k}}^2+2\hat{\tilde{k}}+\Delta)+\alpha e^{-i\varphi_0}\bigg)\nonumber,
\end{align}
where $\hat{\tilde{k}}=-i (d/d\varphi)$. $\varphi$ is measured from the positive direction of the x-axis in anticlockwise sense. For the sake of simplicity we introduced the dimensionless quantity $B+V_1\sin{\varphi}$ which corresponds to the TE-TM splitting, inherited from the plane microcavity, here $B=2\beta m_{\text{eff}}/\hbar^2$ ($\beta$ defines the TE-TM splitting of the plane microcavity without tilt) and $V_1$ accounts for the dependence of the TE-TM splitting on the changing width of the quantum well. $V_0$ describes the position-dependent shift of the energy levels due to the cavity gradient. The dimensionless parameter $\Delta=(\pi R/a)^2$ corresponds to the LT-splitting, stemming from the confinement in the radial direction. We also consider an additional splitting between linear polarizations which acts as a static in-plane field, and described by $\alpha$ and $\varphi_0$. This field is usually linked with the crystallographic axes and appears because of the anisotropy of the quantum well. 


One can relate the parameters $B=\beta m_{\text{eff}}/\hbar^2$ and $V_1$ introduced in Eq.~(\ref{eq:ham_matr}) to the parameters of the DBRs as (see details in Appendix~\ref{Appendix_derivation})
\begin{align}
    &\beta=\frac{2\hbar L_{c,0} L_{\text{DBR}}(\omega_{c,0}-\bar{\omega}) c^2 n_0}{(L_{c,0}+L_{\text{DBR}})^2 n_c^2\omega_{c,0}^2},
\end{align}
and 
 \begin{align}
    &\beta+\frac{\hbar^2V_1}{m_{\text{eff}}}=\frac{2\hbar L_{c,\pi/2} L_{\text{DBR}}(\omega_{c,0}-\bar{\omega}) c^2 n_0}{(L_{c,\pi/2}+L_{\text{DBR}})^2 n_c^2\omega_c^2},
\end{align}
where $n_0$ and $n_c$ are the refractive indexes of the surrounding media and the cavity respectively $L_{c,0}$, $L_{c,\pi/2}$ are the widths of the cavity at the points of the ring defined by $\varphi=0,\pi$, respectively and $\omega_{c,0}$ is the real part of the  cavity eigenfrequency at zero in-plane momentum. $L_{\text{DBR}}=n_a n_b\bar{\lambda}/(2(n_b-n_a))$, which is frequently called the effective length of a Bragg mirror, $n_{a,b}$ are the refractive indices of the layers comprising the DBRs and $\bar{\lambda}$ is the wavelength, corresponding to central frequency of the stop-band $\bar{\lambda}=2\pi c/\bar{\omega}$. In the following discussion, we drop $V_1$ and $\alpha$. The time evolution of the polariton condensate spinor $\Psi$ in presence of energy relaxation and finite lifetime $\tau$ reads:

\begin{equation} \label{eq_nonHermitian_Shroedinger}
    i \hbar {\partial \Psi \over \partial t} = \left( 1 - i \Lambda \right) \hat{H} \Psi -\frac{i \hbar}{2\tau}\Psi
\end{equation}

Energy relaxation is phenomenologically included in the model by multiplying the Hamiltonian \eqref{eq:ham_matr} by the complex coefficient $1 - i \Lambda$, where $\Lambda$ is the dimensionless energy relaxation parameter. This is an energy diminishing scheme, and the higher energy eigenstates relax more quickly than the lower energy eigenstates. At long times of evolution only the ground state survives. We evolve an initial state which is close to the initial state observed in the experiment after hot thermalization. Keeping in mind the left-right mirror symmetry of the ring we pick the initial condensate order parameter with horizontal polarization and localized at the top of the ring,    
\begin{equation}
    \Psi(\varphi,t=0) = \exp\left[ - {(\varphi - \pi/2)^2 \over 2 d^2}  \right] \left[ 1, 1 \right]^\mathrm{T}.
\end{equation}
Numerical solution of Equation \eqref{eq_nonHermitian_Shroedinger} is shown in Fig. \ref{fig:simulation} for two different choices of cavity tilt. It captures the evolution of the polariton spinor after excitation by a single pulse.  

As the initial spin balanced state diffuses from the top, the two components of the spinor are pushed in opposite directions creating a spin imbalanced state. This is shown at early times in the circular polarization components $S_3$ in Fig. \ref{fig:simulation} (d) and (h). In the linear polarization sector, which is a measure of the relative phase between the two spinor components, we find rotation of the polarization as the condensate flows out from the point of generation. This is shown by appearance of non-zero $S_2$ component shortly after evolution of the initial state in Fig. \ref{fig:simulation} (b) and (f). As the condensate fills the entire ring, we see emergence of the four-leaf pattern in $S_1$ and $S_2$. This is clearly seen in Fig. \ref{fig:simulation} (b) and (c) while obfuscated in Fig. \ref{fig:simulation} (f) and (g) due to interference between clockwise and anti-clockwise moving waves. As the condensate settles to the minimum of the trap potential, this pattern is seen at the bottom of the ring in Fig. \ref{fig:simulation} (b), (c), (f) and (g). In this state the circular polarization is nearly absent as the ground state of the ring is not circularly polarized as shown in Appendix \ref{ApdxA}. It should be noted that the model also neglects the evolution of the excitonic reservoir in the ring, which dynamically reshapes the effective potential for polaritons. Evidence of long transport of the reservoir was shown in these rings in Ref. \cite{mukherjee2019observation}. The model introduced in this section captures all the qualitative aspects of the linear and circular polarization precession following a quench, which emphasizes the role of the TE-TM splitting on the condensate dynamics. Finally, we note that the polarization of the condensate observed in the experiment at late times doesn't correspond to any eigenstate of the tilted ring and requires further theoretical investigation into the details of energy relaxation and thermalization processes which could lead to such a pre-thermal state.

\section{Conclusion}\label{Sec5}
Ring shaped polariton waveguides is attracting a great deal of attention for exploring various topological effects \cite{zezyulin2018,kozin2018,zhang2017rotation, KozinYulinPRA2019} due to the TE-TM splitting in these structures. We present in this direction the first experiments on etched polariton rings. We studied the polarization dynamics of a non-equilibrium polariton condensate formed after a pulsed excitation. Through time resolved measurements we were able to observe the temporal signature of intrinsic optical Hall effect shortly after quench. We provide a qualitative estimate of the length scale over which the spatial coherence builds up in the ring by observing spatial energy locking and linewidth narrowing of the emission. We also present a theoretical model which captures qualitatively the formation of a four-leaf pattern in the $S_1$ and $S_2$ components of the Stokes vector, the relative angular phase offset between $S_1$ and $S_2$ and finally the contrasting ratio between the degree of linear and circular polarization in the ring thus elucidating the role of the anisotropic pseudo magnetic field originating from the TE-TM splitting and the tilt in the microcavity structure. 

Future work will explore making the rings radially thinner pushing further apart the radially confined states in energy while also diminishing the cavity tilt. These rings could then serve as an ideal platform for studying one-dimensional macroscopic quantum phenomena similar to superconducting rings. Already with these long lifetime samples we could address interesting questions in nonequilibrium physics, such as generation of long lived non thermal states, which could be observed and studied in this system.     

\section{Acknowledgements}

The work at Pittsburgh was funded by the National Science Foundation (Grant No. DMR-2004570). The work of sample fabrication at Princeton was funded by the Gordon and Betty Moore Foundation (Grant No. GBMF-4420) and by the National Science Foundation MRSEC program through the Princeton Center for Complex Materials (Grant No. DMR-0819860). VKK, AVN and IAS acknowledge support from Russian Science Foundation (project No. 18-72-10110) and Icelandic Science Foundation (Project "Hybrid Polaritonics"). V.K.K  acknowledges the support from the Eimskip Fund. AVN acknowledges support from the European Union’s Horizon 2020 research and innovation programme under the Marie Skłodowska-Curie grant agreement No 846353.

\bibliography{genref}

\appendix

\section{Derivation of Hamiltonian}\label{Appendix_derivation}

	In this section we derive the effective Hamiltonian describing a single polariton ring of radius $R$ and width $a$, accounting for the effect of the TE-TM splitting and the tilt of the well, where only the lowest radial subband is occupied.

	We start with the Hamiltonian of 2D polaritons inside a planar microcavity neglecting the cavity width gradient~\cite{FlayacVortex}  
	\begin{equation}\label{eq:H_2D}
	\widehat{H}_{2D}= 
	\begin{pmatrix}
	\widehat{H}_0(\hat{\mathbf{k}})& \widehat{H}_{\text{TE-TM}}(\hat{\mathbf{k}}) \\
	\widehat{H}^{\dagger}_{\text{TE-TM}}(\hat{\mathbf{k}}) & \widehat{H}_0(\hat{\mathbf{k}})
	\end{pmatrix},
	\end{equation}
	where the diagonal terms $\hat{H}_0$ describe the kinetic energy
	of lower cavity polaritons, and the off-diagonal terms
	$\widehat{H}_{\text{TE-TM}}$ correspond to the TE-TM splitting. We further employ the effective mass approximation  
	\begin{equation}
	\widehat{H}_0(\hat{\mathbf{k}})=\frac{\hbar^2 \hat{\mathbf{k}}^2}{2 m_{\text{eff}}}.
	\end{equation}
	The TE-TM part is given by
	\begin{equation}
	\widehat{H}_{\text{TE-TM}}(\hat{\mathbf{k}})=\beta\left(\frac{\partial}{\partial y}+i\frac{\partial}{\partial x}\right)^2,
	\end{equation}
	where $\beta$ governs the strength of the TE-TM
	splitting and may be expressed in the longitudinal and
	transverse polariton effective masses $m_l$ and $m_t$ as $\beta=(\hbar^2/4)(m_l^{-1}-m_t^{-1})$. In order to proceed with the derivation of the correct 1D Hamiltonian let us pass to the polar coordinates and add the confining potential $V(r)$ confining the polariton wave functions on the ring in the radial direction to the planar cavity Hamiltonian $\widehat{H}_{2D}$. The confining potential is taken as an infinite square well in the radial direction. The terms associated with the TE-TM splitting in polar coordinates read	
	\begin{equation}\label{TETM_cyl}
	\begin{aligned}
	&\left(\frac{\partial}{\partial y}\pm i\frac{\partial}{\partial x}\right)^2=e^{\mp 2 i\varphi}\times\\
	&\left(-\frac{\partial^2}{\partial r^2}\pm \frac{2 i}{r}\frac{\partial^2}{\partial r \partial \varphi}\mp  \frac{2 i}{r^2}\frac{\partial}{\partial\varphi}+\frac{1}{r}\frac{\partial}{\partial r}+\frac{1}{r^2}\frac{\partial^2}{\partial \varphi^2}\right).
	\end{aligned}
	\end{equation}
	We decompose the Hamiltonian Eq.~(\ref{eq:H_2D}) into two parts $\widehat{H}_{2D}=\widehat{H}_0(r)+\widehat{H}_1(r,\varphi)$, where
	\begin{equation}\label{H_0}
	\widehat{H}_0(r)=-\frac{\hbar^2}{2m_{\text{eff}}}\left(\frac{\partial^2}{\partial r^2}+\frac{1}{r}\frac{\partial}{\partial r}\right)+V(r).
	\end{equation}
	Since we can now separate the variables assuming only the lowest radial mode occupation $\widetilde{\Psi}(r,\varphi)=R_0(r)\Psi(\varphi)$, the effective Hamiltonian~\cite{correct1DHam} for $\Psi(\varphi)$ reads
	\begin{equation}\label{eq:Ham_definition}
	\widehat{H}=\langle R_0(r)\lvert\widehat{H}_1(r,\varphi)\rvert R_0(r)\rangle,	
	\end{equation}
    	where $R_0(r)$ is the lowest radial mode of the Hamiltonian (\ref{H_0}). For an infinite square well potential of width $a$ centered at $R$ in the radial direction, the radial modes, which are non-zero only in $[R-a/2,R+a/2]$, are given by 
	\begin{align}
	    &R_n(r)=\\
	    &A_n\bigg(Y_0(\frac{\epsilon_n r}{R})-\frac{Y_0(\epsilon_n(1-\frac{a}{2R}))) _0 F_1(1;-(\frac{\epsilon_n r}{2R})^2)}{_0 F_1(1;-(\epsilon_n (1-\frac{a}{2R}))^2/4)}\bigg),\nonumber
	\end{align}	
	where we introduced the dimensionless energy eigenvalues $\epsilon_n=\frac{\sqrt{2 m E_n}R}{\hbar}$ and $_0 F_1(a;z)$ is the confluent hypergeometric function and $A_n$ is the normalization constant. The eigenvalues $\epsilon_n$ satisfy the following equation
	\begin{align}
	  \frac{Y_0(\epsilon_n (1+\frac{a}{2R})) _0 F_1(1;-(\epsilon_n (1-\frac{a}{2R}))^2/4)}{Y_0(\epsilon_n (1-\frac{a}{2R})) _0 F_1(1;-(\epsilon_n (1+\frac{a}{2R}))^2/4)}=1.
	\end{align}
	Now we perform the averaging over the lowest radial mode $R_0(r)$ corresponding to the lowest eigenvalue $\epsilon_0$. First we observe $\langle R_0(r)\lvert \frac{1}{r}\frac{\partial}{\partial r}\rvert R_0(r)\rangle=\int_0^{\infty}R_0(r)R'(r)dr=R^2_0(r)/2\lvert_0^{\infty}=0$ for any $R_0(r)$ such that $R_0(0)=R_0(+\infty)=0$. Then, we calculate $\langle R_0(r)\lvert \frac{\partial^2}{\partial r^2}\rvert R_0(r)\rangle=-C_0/a^2$, where we used that $R''_0+r^{-1}R'_0=-(2m_{\text{eff}}/\hbar^2) E_0 R_0$ and the fact that from dimensional analysis it follows that $E_0= \hbar^2 C_0/(2 m a^2)$, where $C_0$ is dimensionless. Finally, we have
	\begin{align}
	 &\langle R_0(r)\lvert \frac{\partial^2}{\partial r^2}\rvert R_0(r)\rangle=-C_0/a^2,\nonumber\\ 
	 &\langle R_0(r)\lvert \frac{1}{r}\frac{\partial}{\partial r}\rvert R_0(r)\rangle=0,\nonumber\\ 
	 &\langle R_0(r)\lvert \frac{1}{r^2}\rvert R_0(r)\rangle=\mathcal{F}(a/R)/R^2,
	\end{align}
	where the equation in the last line is obtained from dimensional analysis and $\mathcal{F}(x)$ is a dimensionless function. We approximate $R_0(r)$ by its first Fourier harmonic, which is reasonable as long as only the lowest radial mode is occupied $R_0(r)\approx\sqrt{\frac{2}{aR}}\sin(\frac{\pi(r-(R-a/2)}{a})$. This function is properly normalized $\int_{0}^{\infty} R_0^2(r)rdr=1$ and satisfy the boundary conditions $R_0(R-a/2)=R_0(R+a/2)=0$. Then, we have $C_0=\pi^2$ and for $0<x<1$ it follows $\mathcal{F}(x)\approx 1$ to two decimal places. Thus, we finally arrive at the following hamiltonian
	\begin{align}\label{ring_ham}
	&\widehat{H}=\frac{\hbar^2}{2m_{\text{eff}}R^2}\cdot\nonumber\\ 
	&\begin{pmatrix}   
	\hat{\tilde{k}}^2 & B e^{-2i\varphi}(-\hat{\tilde{k}}^2+2\hat{\tilde{k}}+\Delta) \\
	B e^{2i\varphi}(-\hat{\tilde{k}}^2-2\hat{\tilde{k}}+\Delta) & \hat{\tilde{k}}^2
	\end{pmatrix}
	\end{align}
	where $\hat{\tilde{k}}=-i (d/d\varphi)$ and the energy levels are now shifted by a constant as compared to the Hamiltonian~(\ref{eq:H_2D}).  For the sake of simplicity we introduced a dimensionless parameter $B$ corresponding to the TE-TM splitting as $B=2\beta m_{\text{eff}}/\hbar^2$ and a dimensionless parameter $\Delta=(\pi R/a)^2$ corresponding to the LT-splitting, stemming from the confinement in the radial direction. In the case where the ring is infinitely thin, the model reduces to the one described in~\cite{kozin2018}. 
	
    In general, solutions of the stationary Schrodinger equation with the Hamiltonian~(\ref{ring_ham}) can be represented in the following form
	\begin{equation}\label{eq:ring_eigenfunc}
	\Psi_{k,\alpha}(\varphi)=\widetilde{\chi}_\alpha(\varphi,k) e^{i k R\varphi},
	\end{equation}
	where $\widetilde{\chi}_\alpha(\varphi,k)$ is the corresponding spinor
	\begin{equation}
	\widetilde{\chi}_\alpha(\varphi,k) =\frac{1}{\sqrt{\xi_\alpha(k)^2+1}}\left(
	\begin{array}{c}
	e^{-i\varphi}\\
	\xi_\alpha(k)e^{i\varphi}\\
	\end{array}
	\right),
	\end{equation}
	and 
	\begin{equation}
	\xi_\alpha(k)=-\frac{a^2B(kR-1)(kR-3)+8\pi^2 R^2}{a^2((kR+1)^2-E_k^\alpha)}.
	\end{equation}
	The energy spectrum of the Hamiltonian can be found analytically, the energy levels are given by
		\begin{widetext}
	\begin{equation}\label{eq:energy_dispersion}
	    E_k^{L,U}=1+k^2R^2\mp\frac{\sqrt{B^2\pi^4+2a^2B^2\pi^2(3+k^2R^2)R^{-2}+a^4(4k^2R^2+B^2(9-10k^2R^2+k^4R^4))R^{-4}}}{(a/R)^2}
	\end{equation}
	\end{widetext}
	in $\hbar^2/(2m_{\text{eff}}R^2)$ units.
 As we consider polaritons with spin $\pm1$ as a two-level system, the z-projection of the operator of total angular momentum is $\hat{J}_z=\hbar\hat{\tilde{k}}+\hbar\sigma_z$. One can check that $\hat{J}_z\Psi(\varphi)=\hbar \tilde{k}\Psi(\varphi)$ which clarifies the physical meaning of $\tilde{k}=kR$. 
 The periodic boundary condition imposes the condition $\Psi(\varphi)=\Psi(\varphi+2\pi)$, which yields integer $\tilde{k}$ corresponding to the quantized orbital angular momentum.
 According to Eq.~(\ref{eq:energy_dispersion}) the energy becomes quantized as well. 
	
To account for the tilt of the well one needs to replace $B$ in the Hamiltonian~(\ref{ring_ham}) by $B+V_1\sin{\varphi}$ and add $V_0\sin{\varphi}$ to the diagonal elements, where we assume that the value of the TE-TM splitting as well as the shift of the energy levels linearly depends on the width of the well. Having done that, we arrive at the Hamiltonian~(\ref{eq:ham_matr}), where an additional splitting between linear polarizations is added (see the text) below Eqs.~(\ref{eq:ham_matr}).

The dependence of the TE-TM splitting on the local width of the ring can be calculated as follows. Introducing the coefficient
\begin{equation}
    L_{\text{DBR}}=\frac{n_a n_b\bar{\lambda}}{2(n_b-n_a)},
\end{equation}
which is frequently called the effective length of a Bragg mirror, where $n_{a,b}$ are the refractive indexes of the layers comprising the DBRs and $\bar{\lambda}$ is the wavelength, corresponding to the central frequency of the stop-band $\bar{\lambda}=2\pi c/\bar{\omega}$. Next, introducing
\begin{equation}
    \delta=\omega_c-\bar{\omega}
\end{equation}	
where $\omega_c$ is the real part of the cavity complex eigenfrequency, 
the TE-TM splitting thus reads~\cite{Microcavities2017}
\begin{align}
    &\omega^{\text{TE}}(L_c,\omega_c, \phi_0)-\omega^{\text{TM}}(L_c,\omega_c, \phi_0)\approx\nonumber\\ 
    &\frac{L_c L_{\text{DBR}}}{(L_c+L_{\text{DBR}})^2}\frac{2\cos\phi_{\text{eff}}\sin^2\phi_{\text{eff}}}{1-2\sin^2\phi_{\text{eff}}}\delta,
\end{align}
where $\phi_{\text{eff}}\approx\arcsin{((n_0/n_c)\sin\phi_0)}$, the coefficients and $L_c$ is the width of the cavity, $n_0$ and $n_c$ are the refractive indexes of the surrounding media and the cavity respectively. Now, one can relate the parameters $B=\beta m_{\text{eff}}/\hbar^2$ and $V_1$ introduced in Eq.~(\ref{eq:ham_matr}) as
\begin{align}
    &\beta=\nonumber\\
    &\lim\limits_{k_{||}\rightarrow0} \frac{\hbar( \omega^{\text{TE}}(L_{c,0},\omega_c, \phi_0)-\omega^{\text{TM}}(L_{c,0},\omega_c, \phi_0))}{k_{||}^2}=\nonumber\\
    &=\frac{2\hbar L_{c,0} L_{\text{DBR}}(\omega_{c,0}-\bar{\omega}) c^2 n_0}{(L_{c,0}+L_{\text{DBR}})^2 n_c^2\omega_{c,0}^2},
\end{align}
and 
 \begin{align}
    &\beta+\frac{\hbar^2V_1}{m_{\text{eff}}}=\nonumber\\
    &\lim\limits_{k_{||}\rightarrow0}\frac{\hbar( \omega^{\text{TE}}(L_{c,\pi/2},\omega_c, \phi_0)-\omega^{\text{TM}}(L_{c,\pi/2},\omega_c, \phi_0))}{k_{||}^2}=\nonumber\\
    &=\frac{2\hbar L_{c,\pi/2} L_{\text{DBR}}(\omega_{c,0}-\bar{\omega}) c^2 n_0}{(L_{c,\pi/2}+L_{\text{DBR}})^2 n_c^2\omega_{c,0}^2},
\end{align}
where $k_{||}=(\omega_c/c)\sin\phi_0$ and $L_{c,0}$, $L_{c,\pi/2}$ are the widths of the cavity at the points of the ring defined by $\varphi=0,\pi$, respectively and $\omega_{c,0}$ is a cavity eigenfrequency at $k_{||}=0$, which also depends on the local width of the cavity.

Let us investigate the ground state of a flat ring (no tilt) and in the absence of birefringence. 
For the experimentally relevant values $B=0.052$ and $\Delta=61.36$, the ground state $\Psi_{0,L}$ is non-degenerate and corresponds to $\tilde{k}=0$ of the lower branch (minus sign in Eq.~(\ref{eq:energy_dispersion})), the expression for it reads
	\begin{equation}\label{eq:ground_state}
	\psi_{0,L}=\frac{1}{\sqrt{2}}
	\left(
	\begin{array}{c}
	e^{-i\varphi}\\
	e^{i\varphi}\\
	\end{array}
	\right).
	\end{equation}
The Stokes vector for this state is given by
	\begin{align}
    \vec{S}=\Psi_{0,L}^{\dagger}\vec{\sigma}\Psi_{0,L}=\left(
	\begin{array}{c}
    	-\cos{2\varphi}\\
    	-\sin{2\varphi}\\
    	0
	\end{array}
	\right),
    \end{align}
thus, the ground state of a flat ring with no birefringence is completely linearly polarized with the polarization direction remaining tangential to the ring. It should be noted, that the Stokes vector in the ground state~(\ref{eq:ground_state}) repeats the pattern of the effective magnetic field produced by the TE-TM splitting (shown in Fig.~\ref{fig:Fig6} (b)), but in the XY-plane. We observed a large degree of linear polarization and a small degree of circular polarization in the ring as seen in Fig. \ref{fig:Fig4} and Fig. \ref{fig:simulation}; the TE-TM splitting term in the Hamiltonian mixes the left and right circular components of the pseudo spinor creating a large component of linear polarized state with a small circular component. Although, the four-leaf angular pattern in $S_1$ and $S_2$ are not phase shifted by $\pi/2$ as predicted from the theory but the $2\varphi$ angular dependence of both the patterns point towards the $2\varphi$ angular dependence of the pseudo magnetic field originating from the TE-TM splitting in the ring microcavity. 

\section{Stokes vector measurement}\label{ApdxA}

Jones matrix formalism is a simple method for keeping track of the polarization of light as it interacts with various optical elements. Commonly it is used to characterize only completely polarized light while Mueller matrix formalism can describe a partial polarized state of light. The full polarization state of light is then characterized by a set of four real numbers known as the Stokes vector. In this appendix we discuss our measurement scheme with the help of Jones matrix to characterize any arbitrary state of partially polarized light, so that the connection between our measurements and the state of the polariton spinor remains transparent. 

Our goal is to measure any arbitrary input state $|\psi\rangle$ = $(E_x, E_y)^T$, where $E_x(=|E_x|e^{i \theta_x})$ and $E_y(=|E_y|e^{i \theta_y})$ are complex numbers. Such a state can faithfully represent the state of a completely polarized light. To include partially polarized light we can add to the above state $\epsilon(W_x,W_y)^T$, such that $\langle W_x\rangle = \langle W_y\rangle = 0$ and $\langle W_x^2\rangle = \langle W_y^2\rangle = 1/2$. Therefore, we see that in order to characterize the input state, we need to find just four real numbers $\{|E_x|,|E_y|,\theta_{yx}(=\theta_y-\theta_x),\epsilon\}$ requiring only four measurements summarized in Table \ref{tab:T1}.

The action of $\lambda/2$- and $\lambda/4$-waveplates with fast axis rotated by $\theta$ from the vertical on $|\psi\rangle$ are given by 
\begin{equation}
    \begin{aligned}
    & H(\theta)=R(-\theta)\quad
\begin{pmatrix} 
1 & 0 \\
0 & e^{i\pi}
\end{pmatrix}
\quad R(\theta) \\
& Q(\theta)=R(-\theta)\quad
\begin{pmatrix} 
1 & 0 \\
0 & e^{i\pi/2}
\end{pmatrix}
\quad R(\theta),
 \end{aligned}
\end{equation}
where $R(\theta)$ is the 2D rotation matrix.
\begin{table*}[htbp]
\centering
\begin{tabular}{c c c c c}
\hline
$\theta_{\lambda/2}$ 
& $\theta_{\lambda/4}$ 

& Polarizer orientation & Measured intensity & Observed intensity\\
\hline
0 & $-$  & vertical & $|E_y|^2+\epsilon^2/2$  & $I_1$ \\
\\
$\pi/8$ & $-$  & vertical & $\frac{1}{2}\big(|E_x|^2+|E_y|^2+(E_x^*E_y+E_y^*E_x)+\epsilon^2\big)$ & $I_2$ \\
\\
$\pi/4$ & $-$ & vertical & $|E_x|^2+\epsilon^2/2$ & $I_3$\\
\\
$-$ & $\pi/4$ & vertical & $\frac{1}{2}\big(|E_x|^2+|E_y|^2+i(E_y^*E_x-E_x^*E_y)+\epsilon^2\big)$ & $I_4$\\
\\
$-$ & $7\pi/4$ & vertical & $\frac{1}{2}\big(|E_x|^2+|E_y|^2-i(E_y^*E_x-E_x^*E_y)+\epsilon^2\big)$ & $I_5$\\
\hline
\end{tabular}
\caption{}
\label{tab:T1}
\end{table*}
Since $\lambda/2$- and $\lambda/4$-waveplates have different thicknesses, so we expect them to have slightly different transmission efficiency. To compensate for this we found that it is usually more accurate to take two measurements with $\lambda/4$-waveplate which could be added to give the total intensity of light $I_{tot}$ ($=S_0$). From our measurements we can calculate the components of the Stokes vector by taking linear combinations as shown below.
\begin{equation}
\begin{aligned}
    & I_{tot}=I_1+I_3=I_4+I_5=|E_x|^2+|E_y|^2+\epsilon^2\\
    & S_1 = \frac{I_3 - I_1}{I_1+I_3} = \frac{|E_x|^2-|E_y|^2}{|E_x|^2+|E_y|^2+\epsilon^2}\\
    & S_2 = \frac{2I_2 - I_1-I_3}{I_1+I_3} = \frac{E_x^*E_y+E_y^*E_x}{|E_x|^2+|E_y|^2+\epsilon^2}\\
    & S_3 =\frac{I_5 - I_4}{I_4+I_5} = \frac{i(E_x^*E_y-E_y^*E_x)}{|E_x|^2+|E_y|^2+\epsilon^2}\\
    \end{aligned}
\end{equation}

It is also straightforward to see that the intensity of the polarized and unpolarized light are given by
\begin{equation}
\begin{aligned}
& I_{pol} = |E_x|^2+|E_y|^2 = I_{tot}\sqrt{S_1^2+S_2^2+S_3^2}\\
&    \epsilon^2=I_{tot}(1-\sqrt{S_1^2+S_2^2+S_3^2}).
    \end{aligned}
\end{equation}
The relative phase difference between the spinor component is 
\begin{equation}
    \tan\theta_{yx} = \frac{S_3}{S_2}
\end{equation}
and the amplitudes are given by,
\begin{equation}
\begin{aligned}
  &  |E_x|^2 = \frac{I_{tot}}{2}\Big(\sqrt{S_1^2+S_2^2+S_3^2}+S_1\Big) \\
  &  |E_y|^2 = \frac{I_{tot}}{2}\Big(\sqrt{S_1^2+S_2^2+S_3^2}-S_1\Big).
    \end{aligned}
\end{equation}
The direction of the polarization plane $\varphi$ is given by,
\begin{equation}
\varphi = \frac{1}{2}\tan^{-1}\frac{S_2}{S_1}.
\end{equation}

\end{document}